\documentstyle[pre,aps,epsfig]{revtex}
\begin{document}
\draft

\twocolumn[\hsize\textwidth\columnwidth\hsize\csname@twocolumnfalse\endcsname
\title{A series representation of the nonlinear equation for axisymmetrical fluid membrane shape}
\author{B. Hu$^{1,2}$, Q. H. Liu$^{1,3,4}$, J. X. Liu$^{4}$, 
X. Wang$^{2}$, H. Zhang$^{1}$, O. Y. Zhong-Can$^{4}$}
\address{$^{1}$ Department of Physics and the Center for Nonlinear Studies, Hong 
Kong Baptist University, Hong Kong, China}
\address{$^{2}$ Department of Physics, University of Houston, Houston,TX77204-5506,
USA}
\address{$^{3}$ Department of Applied Physics, Hunan University, Changsha, 410082,
China}
\address{$^{4}$ Institute of Theoretical Physics, Academia Sinica, P.O. Box 2735, 
Beijing, 100080, China}
\date{\today}
\maketitle

\begin{abstract}
Whatever the fluid lipid vesicle is modeled as the spontaneous-curvature,
bilayer-coupling, or the area-difference elasticity, and no matter whether a
pulling axial force applied at the vesicle poles or not, a universal shape
equation presents when the shape has both axisymmetry and up-down symmetry.
This equation is a second order nonlinear ordinary differential equation
about the sine $sin\psi (r)$ of the angle $\psi (r)$ between the tangent of
the contour and the radial axis $r$. However, analytically there is not a
generally applicable method to solve it, while numerically the angle $\psi
(0)$ can not be obtained unless by tricky extrapolation for $r=0$ is a
singular point of the equation. We report an infinite series representation
of the equation, in which the known solutions are some special cases, and a
new family of shapes related to the membrane microtubule formation, in which 
$sin\psi (0)$ takes values from $0$ to $\pi /2$, is given.
\end{abstract}


\preprint{Submit to PRE}
\vskip2pc]

%%%%%%%%%%%%%%%%%%%%%%%%%%%%%%%%%%%%%%%%%%%%%%%%%%%%%%%%%%%%%%%%%%%%%%
%  Text   %%%%%%%%%%%%%%%%%%%%%%%%%%%%%%%%%%%%%%%%%%%%%%%%%%%%%%%%%%%%
%%%%%%%%%%%%%%%%%%%%%%%%%%%%%%%%%%%%%%%%%%%%%%%%%%%%%%%%%%%%%%%%%%%%%%

Vesicles are closed surfaces of amphiphilic molecules dissolved in water,
which form flexible bilayer membranes in order to minimize contact between
the hydrocarbon chains of the lipid and water\cite{lip}. Recently, the
vesicle shapes have attracted wide interest from different communities such
as physics\cite{lip,ben,oyliu,seif1}, mathematics\cite{nit,oyliu,ben},
chemistry\cite{gra,eri}, and biology\cite{seif1,hel0}. The naive model was
given by Canham\cite{can}, in which only the surface bending elasticity is
considered. An important progress was the introduction of the spontaneous
curvature into the Canham's theory\cite{hel}, and it was made by Helfrich
with analogue to the spontaneous splay in the liquid crystal molecular
layer. Since the membrane consists in two monolayers, if assuming that two
layers depart from each other at a fixed distance\cite{shsi,evan}, a
so-called bilayer-coupling model was explored\cite{comb1}. However, the
individual monolayers can in fact expand elastically under tensile stress,
the so-called area-difference elasticity model \cite{comb1} was introduced
and investigated\cite{miao1,dob,ben}. The latter three models are more
realistic than the naive one, and have been intensively studied in recent
years. A surprising common property of these three models is that they give
exactly the same shapes\cite{seif2,sve6,miao1}, while differing from each
other in accounting for the shape transitions, such as budding and
vesiculation transition, transition from prolates to oblates etc. \cite
{seif2,miao2}. Furthermore, pulling or pushing the vesicle axially making it
up-down and axis symmetrical, as it used in the experimental investigation
of the tether formation\cite{hotani,sve6}, does not complicate the form of
the mathematical equation for the general axisymmetrical vesicle in the
spontaneous curvature model without force included. But the meanings of
parameters need to re-specify. Except some numerical approach using the
powerful software Surface Evolver\cite{yan}, usual numerical as well as
analytical method is employed to solve the shape equation representing
axisymmetrical vesicle. However, in analytical side\cite{comb2,oy1}, there
is \ not a general applicable method to study the equation. In numerical
side, $r=0$ (cf. \ the following Eq.(1) ) is a singular point of the
equation, so the shape around the point cannot be obtained unless by tricky
extrapolation\cite{seif2}. In this paper, we are going to reveal a novel
property of the equation: the equation has an equivalent series
representation, in which the two above problems do not present anymore. As a
meaningful demonstration, a new family of shape related to the tether
formation will be given.

Since the detailed or key steps of the derivation of the same equation from
different models is available in many papers, e.g.\cite{sve6,zhe,liu3}, we
can start our further discussion from the equation itself with clearly
specifying the physics meaning of each quantity in it. This may of course
makes this paper more concise. And the equation we start from reads: \cite
{zhe,sve6,liu3}

\begin{eqnarray}
&&cos^{2}\psi \frac{d^{2}\psi }{dr^{2}}-\frac{sin(2\psi )}{4}(\frac{d\psi }{%
dr})^{2}+\frac{cos^{2}\psi }{r}\frac{d\psi }{dr}  \nonumber \\
&&-\frac{sin(2\psi )}{2r^{2}}-\frac{\delta pr}{2kcos\psi } \nonumber \\
&&-\frac{sin\psi }{2cos\psi }(\frac{sin\psi }{r}-c_{0})^{2}-\frac{\lambda
sin\psi }{kcos\psi }=\frac{C}{rcos\psi }.  \label{zheliu}
\end{eqnarray}
The quantities in this equation need some explanations. For an
axisymmetrical shape, we can choose the symmetrical axis to be the $z$-axis.
The contour of the shape can be plotted in $rz$ plane, with $r$ being the
radial coordinate denoting the distance from the symmetric $z$ axis. Then
the tangent angle $\psi (r)$ of the contour is measured clockwise from $r$
axis. Using $s$ to denote the arclength along the contour measured from the
north pole of the shape and the $\phi $ the azimuthal angle, we have the
following geometrical relations: 
\begin{eqnarray}
&&\left\{ 
\begin{array}{l}
\frac{dr}{ds}=cos\psi (r) \\ 
\frac{dz}{ds}=-sin\psi (r) \\ 
z(r)-z(0)=-\int_{0}^{r}tan\psi (r)dr
\end{array}
\right. \\
&&{\bf n}=(sin\psi cos\phi ,sin\psi sin\phi ,cos\psi )
\end{eqnarray}
in which ${\bf n}$ denotes the normal of the surface. As to constants $k$, $%
C $, $c_{0}$, $\delta p$, and $\lambda $, they have different meaning in
different models. Taking the spontaneous curvature theory as example, $k$
the elastic modulus, $C$ the integral constant, $c_{0}$ the spontaneous
curvature, $\delta p$ and $\lambda $ the two Lagrangian multipliers taking
account for the constraints of constant volume and area, which may be
physically understood as the osmotic pressure between the ambient and the
internal environments, and the surface tension coefficient, respectively.
When $C=0$, Eq. (\ref{zheliu}) reduces to be the original Helfrich shape
equation\cite{zhe,deu}. When $c_{0}\rightarrow N$, $(2\lambda
+c_{0}^{2})/4\rightarrow -L$, $\delta p\rightarrow -6M$, $C\rightarrow -4F$,
Eq.(1) becomes the equation representing the shapes with the same symmetry
in area-difference elasticity model with the inclusion of the axially
external force $F$ (Eq. (19) in \cite{sve6}).

Although no one attempts to analytically solve it in a systematical way, we
know that the following simple functions.

\begin{equation}
sin\psi (r)=ar+b/r+c,~~\text{~and},~~~ar+c_{0}rlnr.
\end{equation}
solve the equation\cite{oy1}. Adjusting the constants $a$, $b$, and $c$ in
permissible ranges, the former form of the solution gives following shapes:
1) cylinder $r=r_{0}$, $sin\psi (r_{0})=1$; 2) the axisymmetrical constant
mean curvature surfaces ( the so-called the Delaunay surfaces) with $c=0$
and an extension with $c\not=0$; and 3) Clifford torus: $sin\psi (r)=-\sqrt{2%
}+r$. The latter form of solution gives shapes from the self-intersecting
oblates, oblate, sphere, prolate to very long capped prolate\cite{liu2}.
These analytical solutions not only have explained the existing but puzzling
shapes\cite{oyliu,liu3}, but also had predicted something new. Clifford
torus, for instance, was first theoretically predicted, then experimentally
confirmed\cite{bensimon}.

We will show that Eq.(\ref{zheliu}) can be analytically solved for the
physically interesting case. For convenience, we make a transformation for
Eq. (\ref{zheliu})

\begin{equation}
\psi(r)=arcsinf(r)
\end{equation}
and rescaling the unit such that $k=1$. Then we get from Eq.(\ref{zheliu})
an equation for $f(r)$:

\begin{eqnarray}
&&-2Cr-\delta pr^{3}-2f(r)-c_{0}^{2}r^{2}f(r)-2\lambda r^{2}f(r)  \nonumber
\\
&&+2c_{0}rf^{2}(r)+f^{3}(r)+2krf^{\prime }(r)-2rf^{2}(r)f^{\prime }(r) 
\nonumber \\
&&+r^{2}f(r){f^{\prime }}^{2}(r)+2r^{2}f^{\prime \prime
}(r)-2r^{2}f^{2}(r)f^{\prime \prime }(r)=0  \label{diff}
\end{eqnarray}
This is a second order nonlinear ordinary differential equation (ODE) not
belonging to any well-studied mathematical type. For our purpose to attack
it, let us recall the Taylor series expansion of a function. As well-known,
for any analytical function $f(r)$ defining in a closed interval $r\in
\lbrack r_{1},r_{2}]$, global Taylor expansion of $f(r)$ around a point $%
r_{0}$ is 
\begin{eqnarray}
f(r) &=&\sum_{k=0}^{n}\frac{f^{(k)}(r_{0})(r-r_{0})^{k}}{k!}+R_{n}(r-r_{0}) 
\nonumber \\
&=&\sum_{k=0}^{\infty }\frac{f^{(k)}(r_{0})(r-r_{0})^{k}}{k!},
\label{taylor}
\end{eqnarray}
where $R_{n}(r-r_{0})=(r-r_{0})^{(n+1)}f^{(n+1)}(\xi )$ with $r_{1}<\xi
<r_{2}$. Both the above series are exact and equivalent to each other. The
first finite series is easy to serve the digit purpose, while the second
infinite one is useful to obtain a closed expression once the general form
is determined. To note that every axisymmetrical shape must intersect the
equatorial plane at right angle $\psi =\pi /2$. This means that there exits
a point $r_{0}$, such that $f(r_{0})=1$. At this point $r_{0}$, the two
terms containing the second derivative in Eq.(\ref{diff}) cancel. We have
then value of its first derivative $f^{\prime }(r_{0})$ of $f(r)$ at point $%
r_{0}$ as 
\begin{equation}
f^{\prime }(r_{0})=\pm \frac{\sqrt{%
1+2Cr_{0}-2c_{0}r_{0}+c_{0}^{2}r_{0}^{2}+2\lambda r_{0}^{2}+\delta pr_{0}^{3}%
}}{r_{0}}.  \label{deri}
\end{equation}
Since $f(r)\leq f(r_{0})=1$, we have $f^{\prime }(r_{0})=%
%TCIMACRO{\underset{r\rightarrow r_{0}}{\lim }}%
%BeginExpansion
\mathrel{\mathop{\lim }\limits_{r\rightarrow r_{0}}}%
%EndExpansion
(f(r)-1)/(r-r_{0})\geq 0$ if and only if $r_{0}$ is the maximum radius of
vesicle. Since we treat $f(r)$ a single valued function of $r$ and $r_{0}$
is really the maximum radius of vesicle, only the positive $f^{\prime
}(r_{0})$ is relevant. To note that the physically interesting vesicles are
either free or axially forced, and there is no force or torque acting on the
waist. It implies that the high rank derivatives of \ $f(r)$ at $r_{0}$ as $%
f^{(k)}(r_{0}),k=2,3...$ \ exist, and they can be obtained by the following
method. Taking derivative with respect to variable $r$ in both side, Eq.(\ref
{diff}) becomes a third order differential equation. Putting $r=r_{0}$ and
substituting the value $f(r_{0})$ $(=1)$ and $f^{\prime }(r_{0})$ (given by
Eq. (\ref{deri}) ) into the third order equation, the third order terms
cancel, and we can obtain the value of its second derivative $f^{\prime
\prime }(r_{0})$. Similarly, taking derivative with respect to variable $r$
in both side of the third order derivative equation, we can obtain $%
f^{(3)}(r_{0})$, and so forth, all its higher rank derivatives $%
f^{(k)}(r_{0}),k=4,5...$ can be determined. It is a remarkable result: the
nonlinear ODE (\ref{diff}) can be transformed into an infinite series with
the recurrence relation between coefficients uniquely determined by the
equation itself\cite{footnote}. In the mathematical point of view, all its
solutions with physical significance are obtained.

In this paper, we only treat the simplest aspect of the solution without
considering the congruence of the segments of shapes. Since the
parameterization of the axisymmetrical surface is $\psi (r)$ rather than $%
\psi (s)$, from Eq.(1) $f(r)$ is a single value function of $r$ in interval $%
\psi (r)\in \lbrack 0,\pi /2]$. We can therefore expand $f(r)$ in Taylor
series directly. Conversely, once all Taylor coefficients $%
f^{(k)}(r_{0})/k!,(k=1,2,3,...)$, are known, the function $f(r)$ is
determined from the relation (\ref{taylor}). In principle, we can give such
a series for each shape. As checks of our method, we like to give two
examples. 1) Cylinder with arbitrary radius $r=r_{0}$. It is the case $%
f(r_{0})\equiv 1$ provided all parameters $k,\delta p,\lambda ,c_{0},C$
satisfying $\delta pr_{0}^{3}+2\lambda r_{0}^{2}+k(c_{0}r_{0}-1)^{2}=0$ and $%
C=0$\cite{zhe}. It is the result discussed previously\cite{oy2,zhe}. 2) The
Delaunay surfaces $f(r)=a r+b/r$ (a, b are two constants). These constant
mean curvature surfaces appear when $\lambda =\delta p=C=0$. Now we have
simply $f^{(1)}(r_{0})=\pm (c_{0}-1/r_{0})$ at position $r_{0}$ satisfying $%
f(r_{0})=1$. We study the case with positive sign. We can easily find $%
f^{(k)}(r_{0})=(-1)^{(k+1)}k!(c_{0}r_{0}-2)/(2r_{0}^{k})$, $(k=2,3,4...)$.
Then the solution can be exactly written as $f(r)=a r+b/r$ with $a=c_{0}/2$, 
$b=(r_{0}-c_{0}r_{0}^{2}/2)$. It is the result discussed also previously\cite
{oy1}. One can verify this method using another analytical or numerical
solution as exercises, and can easily find that they are some special cases
of the infinite series with special relation between the parameters.

Now we wish to give new family of shapes. These shapes can be inferred from
the works of three research groups in studying the meaning of the integral
constant $C$. First by Zheng and Liu\cite{zhe}, they noted that the constant 
$C$ roughly meant that the constant pressure difference $\delta p$ in
Helfrich shape equation\cite{zhe} was replaced by $\delta p+2kC/r^{2}$.
Since that the all Helfrich shapes with axisymmetry are smooth at the poles $%
r=0$\cite{deu,seif2}, the constant $C$ means a stress singularity at two
poles. This singularity may lead to two horns at both poles. Secondly,
J\"{u}licher and Seifert claimed that the nonvanishing $C$ connected to
shapes with torus topology\cite{seif3}. Thirdly, Podgornik, Svetina, and \u{Z%
}ek\u{s} in \cite{sve5} and Bo\u{z}i\u{c}, Svetina, and \u{Z}ek\u{s} in \cite
{sve6} showed that a point axial force can lead to such a constant $C$\cite
{sve5,sve6}. We have known that such a force can lead to microtubule
formation of vesicle \cite{sve6,hotani}, and in real experiment a finite
force is enough to produce such structure. However, the expected shapes with
two horns at both poles as exerting finite force have not been obtained
theoretically.

For our purpose, we treat a case in which the quantity in the square of Eq.(%
\ref{deri}) to form the square of a quantity with nonzero $C$. For our
purpose, we choose the simplest parameters as $C=c_{0}$, $\delta p=0,\lambda
=-c^{2}/2$, and the positive sign of $f^{\prime }(r_{0})$. The first seven
derivatives $f^{(k)}(r_{0})$, $(k=0,1,2,..6)$ are

\begin{eqnarray}
&&\{1,{\frac{1}{r_{0}}},{\frac{2\,C}{r_{0}}},{\frac{-4\,C}{{r_{0}^{2}}}},{%
\frac{2\,\left( 35\,C+4\,{C^{2}}\,r_{0}\right) }{5\,{r_{0}^{3}}}},  \nonumber
\\
&&{\frac{-2\,C\,\left( 1155+268\,C\,r_{0}+12\,{C^{2}}\,{r_{0}^{2}}\right) }{%
35\,{r_{0}^{4}}}},  \nonumber \\
&&{\frac{8\,C\,\left( 5040+1689\,C\,r_{0}+99\,{C^{2}}\,{r_{0}^{2}}+32\,{C^{3}%
}\,{r_{0}^{3}}\right) }{105\,{r_{0}^{5}}}}\}.
\end{eqnarray}

Since the conformal invariance of the shape\cite{lip,seif1,liu2,sve6}, the
magnitude of $C$ does not matter. We can therefore choose $C=1$. Then the
Taylor series Eq.(\ref{taylor}) can not give real and positive root of $%
r_{0} $ for equation $f(0)$= a negative number. It means that in this case
the equation is not self-consistent. In contrast, it can give a unique and
positive root for equation $f(0)$= a positive constant. To note that $%
arcsinf(0)$ is the angle between the tangent of the contour at pole and the $%
r$ axis. For a sequence of numbers of $f(0)$ as

\begin{equation}
N=\{0,0.1,0.2,0.3,0.4,0.5,0.6,0.7,0.8,0.9,0.972,1\},
\end{equation}
we have form the equation $f(0)= N$ a corresponding sequence of unique and
(semi-)positive roots for $r_0$ as

\begin{eqnarray}
r_{0} &=&\{0,0.0095195,0.0189412,0.0282679,0.0375022,  \nonumber \\
&&0.0466463,0.0557024,0.0646727,0.0735589,  \nonumber \\
&&0.0823627,0.0886836,0.0910858\}.
\end{eqnarray}
Do not care the magnitude of these values. The energy of the vesicle depends
on the shape due to the conformal invariance. For clearly specifying each
shape, a conformal invariant, the so called relative volume $v=V/(4\pi
R^{3}/3)$ with $R=\sqrt{A/4\pi }$ is usually used\cite{seif1}, where $V$ and 
$A$ are the volume and area of the vesicle respectively. The one-to-one
corresponding relative volumes are:

\begin{eqnarray}
v &=&\{1,0.999971,0.999879,0.999725,0.999505,  \nonumber \\
&&0.999213,0.998844,0.998386,0.997819,  \nonumber \\
&&0.997107,0.996449,0.996136\}
\end{eqnarray}

In FIG. 1, the two limit shapes, sphere (thick solid line) and the prolate
with the sharpest horn (thin line), are plotted. Other shapes with the horns
of angles $\psi(0)$ satisfying $0<\psi(0)<\pi/2$ will occupy the domain
between these two.

For studying the relation between the ``external force'' $C$ and the
relative volume, we define a dimensionless external force as $C(z_{0}-r_{0})$
with $z_{0}$ being half of the length of the prolate, and rescale the
relative volume as $v^{\prime }=4.725(1-v)$. We can find that both the
dimensionless external force and the rescaled relative volume have nearly
the same relationship to the zenith angle, as shown in FIG. 2. Results
plotted in FIG.1 and FIG.2 clearly show that the zenith angle becomes
sharper as the external force becomes larger, or equivalent, the less the
relative volume, the sharper the horn.

\vspace*{1cm} 
\begin{figure}
\centerline{\psfig{file=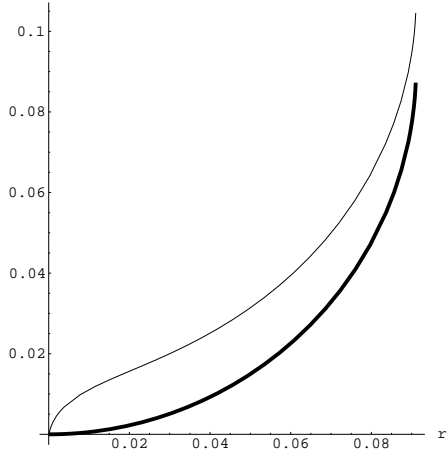,height=6cm}}
\vspace*{1.0cm} 
\caption{The two limit shapes: sphere (thick solid line) and the prolate
with right angle horns on two poles (thin solid line). They have relative
volumes $1$ and $0.996$ respectively. Only a quarter of the contour is
plotted.}
\label{fig1}
\end{figure}

\vspace*{1cm} 
\begin{figure}
\centerline{\psfig{file=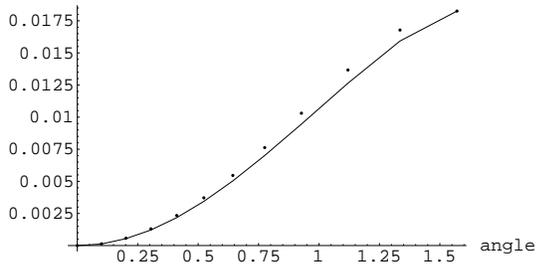,height=3.5cm}}
\vspace*{1.0cm} 
\caption{The correlation between the axial force $C$ and the relative volume 
$v$. We plot the dimensionless force $C(z_{0}-r_{0})$ (solid line) with the $%
z_{0}$ being half of the length of the prolate and the rescaled relative
volume $v^{\prime }=4.725(1-v)$ (solid dots). The larger axial external
force and the smaller relative volume, the sharper angle of the corn.}
\label{fig2}
\end{figure}

In summary, we have found that the second order nonlinear different equation
for axisymmetrical fluid vesicle shapes with up-down symmetry has an
equivalent Taylor series representation, in which the recurrence relation of
Taylor coefficients are uniquely and completely determined be the equation.
Using this representation, we can not only reproduce the known solutions,
but also new ones. The new solutions discussed in this article give the
shapes with horns at two poles. These horns can never form simultaneously if
no axially external force exerted. So, our study directly answers a question
whether the constant $C$ is related to the presence of the axially external
force. This result is compatible with that obtained previously\cite
{sve5,sve6,seif3}. The obtained shapes are related to the formation of the
membrane microtubule on the axially strained vesicles, and finite force is
enough to lead to such structure. Results also show that the shapes with
less relative volume is easier to form such microtubule.

Finally, we give two comments. First, even the Taylor series method is
useful in obtaining and analyzing the solution to the equation, one should
be careful to check the convergence of the series. In fact, we explore many
cases with different parameter set of $\delta p, \lambda, c_0, C$, the
convergence is sometime uncertain. Secondly, the presence of the constant $C$
may not be the sufficient condition to relate the axial force\cite
{sve6,liu2,hotani}, especially in the torus topology situation\cite{seif3}.

{\bf Acknowledgments}

One of the authors (Liu) is indebted to Profs. Peng Huan-Wu, Vipin
Srivastava and Dr. ZhouHaijun for enlightening discussions. This subject is
supported in part by grants from the Hong Kong Research Grants Council
(RGC), the Hong Kong Baptist University Faculty Research Grant (FRG), and in
part by National Natural Science Foundation of China.

%%%%%%%%%%%%%%%%%%%%%%%%%%%%%%%%%%%%%%%%%%%%%%%%%%%%%%%%%%%%%%%%%%%%%
%  References  %%%%%%%%%%%%%%%%%%%%%%%%%%%%%%%%%%%%%%%%%%%%%%%%%%%%%%
%%%%%%%%%%%%%%%%%%%%%%%%%%%%%%%%%%%%%%%%%%%%%%%%%%%%%%%%%%%%%%%%%%%%%


\begin{references}
\bibitem{footnote}  As far as we know, this method offers a new mean to
attack a nonlinear ODE in general. The successful examples will be presented
elsewhere.
\end{references}
\end{document}